\documentclass[%
 reprint,
 amsmath,amssymb,
 aps,
 prfluids,
]{revtex4-2}

\usepackage{graphicx}% Include figure files
\usepackage{dcolumn}% Align table columns on decimal point
\usepackage{bm}% bold math
\usepackage{amsfonts}
\usepackage{amsmath}
\usepackage{xcolor}
\usepackage{color}
\usepackage{ulem}

\begin{document}

\preprint{APS/123-QED}

\title{Exploring 2D turbulent properties in anisotropic and disordered Fourier space: Insights into inverse cascades and universal superdiffusion from randomly sampled triadic interaction
}

\author{Francesco Carbone}
\email{francesco.carbone@cnr.it}
\affiliation{National Research Council, Institute of Atmospheric Pollution Research, University of Calabria, 87036 Rende, Italy.}

\author{Sergio Servidio}%
\affiliation{Dipartimento di Fisica, University of Calabria, 87036 Rende, Italy.}%

\date{\today}% It is always \today, today,
             %  but any date may be explicitly specified

\begin{abstract}
	Two-dimensional turbulent properties are investigated within a ``low-density'' Galerkin-truncated system, with a focus on both Eulerian and Lagrangian characteristics. 
	In particular, an ordered pseudo-logarithmic and a disordered distribution of {active (i.e. resonant)} triads has been sampled in Fourier space, allowing for a tunable degree of anisotropy and ``triadic density'', enabling investigation into their effects on the inverse energy cascade and particle pairs diffusion.	
	Despite the non-uniform and anisotropic mesh in the Fourier 
	space, this reduced model successfully captures 2D turbulence scaling laws and maintains integral energy cascade properties. It consistently reveals the classical double-cascade: a $k^{-5/3}$ inverse cascade at large scales and a $k^{-3}$ direct cascade at small scales, observed across all configurations. Furthermore, while anisotropy, controlled via angular sampling, significantly impacts the vorticity field organization and the efficiency of the inverse energy flux, the system's diffusive properties exhibit a Richardson superdiffusive scaling, $\ell^2(t)\sim t^3$, for particle pair separation. The prescribed spectral anisotropy affects the Lagrangian eddy diffusivity, enhancing diffusion along one direction for short timescales. Conversely, for longer times, particles become uncorrelated, and the separation distance degenerates into the classical Brownian scaling, $\ell^2(t)\sim t$.
	The observed $t^3$ pair-dispersion indicates that the retained spectral interactions sustain super-ballistic separation, while anisotropy mainly affects the dispersion amplitude without modifying the scaling.
\end{abstract}

%\keywords{Suggested keywords}%Use showkeys class option if keyword
                              %display desired
\maketitle

%\tableofcontents

\section{Introduction}

The numerical investigation of fluid turbulence presents formidable challenges due to the vast range of active scales and the inherent nonlinearity of the governing Navier-Stokes (NS) equations. Traditional, Direct Numerical Simulations {(DNS)} on uniform Fourier grids become computationally prohibitive for very high Reynolds numbers ($\textnormal{Re}$), where energy cascades span many orders of magnitude, and the number of degree of freedom ($N_{\textnormal{DOF}}$) required to resolve the entire spectrum scales as a power of $\textnormal{Re}$~\cite{Frish1995,Tran2009}.
This limitation spurred the development of more efficient spectral representations capable of capturing essential physics with reduced computational cost and addressing fundamental questions in fluid dynamics. 

Non-uniform lattices offer a compelling discretization strategy, for example by logarithmically concentrating wave modes in the Fourier space~\cite{Campolina2021}. 
This approach enabled high-resolution two-dimensional (2D) Euler simulations, demonstrating stable, long-term vortex evolution and providing a robust framework for investigating finite-time singularities in inviscid flows. When extended to three-dimensional reversible Navier-Stokes equations~\cite{Gallavotti1996,Costa2023}, this methodology confirmed a second-order phase transition from laminar to turbulent states, with critical exponents consistent with thermodynamic analogies.

Alongside logarithmic lattices, Fourier decimation techniques~\cite{Frisch2012} enhance computational efficiency by retaining a subset of Fourier modes. This approach enables the investigation of turbulent properties in non-integer effective dimensions ($D$).
A profound result is the theoretical prediction and numerical verification of an equilibrium Gibbs state ($k^{-5/3}$ spectrum) at a critical dimension $D_c=4/3$, linking turbulent dynamics to statistical mechanics.
Furthermore, the inverse energy cascade~\cite{Boffetta2011} persists even for $D < 2$, with the Kolmogorov constant diverging as $D \to 4/3$, offering critical insights into energy transfer in constrained or irregular Fourier spaces. Building on these findings, fractal decimation's impact on Lagrangian statistics was explored~\cite{Buzzicotti2016}, revealing the crucial role of active mode distribution. However, {it has been shown that} fractal decimation leads to stronger intermittency and more pronounced anomalous scaling in Lagrangian velocity increments compared to homogeneous decimation, underscoring how Fourier space structure fundamentally influences real-space turbulent transport and mixing.

Here, a different approach is proposed that synergistically combines elements of logarithmic lattices and decimation. A Galerkin truncation of the non-dimensional Navier-Stokes (NS) equations for an incompressible fluid on a two-dimensional torus ($0 \leq (x,y) \leq L \equiv 2\pi$) is performed using a large, but finite, number of Fourier modes distributed in a pseudo-logarithmic and randomized manner.
This specific configuration allows for the observation of key 2D turbulence features, such as a direct enstrophy cascade and an inverse energy flux towards large scales. 

The random truncated Galerkin approach should be distinguished from previous DNS-based studies employing logarithmically concentrated Fourier grids or Fourier decimation techniques. 
In those methods, the Navier--Stokes equations are evolved either in physical space or in a mixed spectral--physical formulation, so that nonlinear interactions formally involve the full convolution over all dynamically accessible modes. 
In practice, nonlinear terms are evaluated in physical space and then transformed back to Fourier space. As a consequence, nonlinear interactions populate the entire resolved spectral domain. Even when only a subset of Fourier modes is retained or emphasized, repeated transformations between physical and Fourier space lead to unavoidable spectral mixing, requiring projection, filtering, or decimation procedures to enforce the desired truncation.

Here, the reduction is implemented directly at the level of the governing equations through a Galerkin projection onto a prescribed finite set of Fourier modes. 
Only triads formed by retained wave vectors are included in the nonlinear interactions, while all other couplings are excluded by construction. 
The resulting system therefore constitutes a closed finite-dimensional dynamical system, and no spectral repopulation or subsequent filtering is required during time integration.

{
For this reason, the method does not aim at approximating a full DNS through a reduced representation, but rather at investigating the intrinsic dynamics of a spectrally truncated system defined {a priori}. 
While certain statistical features may resemble those observed in DNS, the underlying dynamical construction is fundamentally different. Accordingly, the reduced representation is assessed a posteriori through the statistical behavior generated by the retained nonlinear interactions, including inverse energy transfer, spectral properties, and Lagrangian pair-dispersion properties.
}

In addition, the explicit control over the retained wave vector set allows the geometry of active interactions to be prescribed, enabling systematic investigations of anisotropy and interaction topology that cannot be directly imposed in standard DNS--based reductions.
This flexibility is especially relevant for the study of transport and diffusive processes, as the geometry of the active spectral interactions can strongly influence particle dispersion, mixing properties, and energy transfer pathways. The possibility of directly controlling anisotropy and interaction topology therefore represents a distinctive feature of the present Galerkin-truncated formulation.

In the reduced formulation, anisotropy can be prescribed directly at the level of the governing equations by selecting {a priori} the set of retained Fourier modes, including the deliberate introduction of excluded or ``forbidden-interaction'' regions in spectral space. In this way, the anisotropic structure is incorporated in the dynamical system itself through the explicit specification of which wave vectors and triads are allowed to interact.

Such a spectral construction provides an idealized framework capable of mimicking situations in which anisotropy emerges in the full Navier--Stokes dynamics due to mechanisms such as selective damping, large-scale gradients, or externally imposed anisotropic forcing.

Since atmospheric and oceanic flows can be approximated as two-dimensional on a geostrophic scale, the study of 2D turbulence with a prescribed degree of anisotropy is crucial for investigating how flow dynamics could promote the manifestation and intensification of extreme weather events, such as hurricanes, typhoons, and atmospheric rivers~\cite{Ghil2010,Stoll2020,Gencarelli2026}. This is because external factors (e.g., terrain orography, large-scale shear, updrafts due to evaporation and latent heat release) alter flow properties by modifying specific interactions in Fourier space. This directly governs the transfer of energy, which is a principal aspect of these phenomena's dynamics.

As customary, the dynamics of the flow are described by the momentum equations, projected onto the axes of the orthogonal coordinate system, coupled with the continuity equation:
\begin{eqnarray}
\frac{\partial \mathbf{u}}{\partial t} +(\mathbf{u}\cdot\nabla)\mathbf{u} &=& -\nabla P + \nu\nabla^2\mathbf{u} -\lambda \mathbf{u} + \mathbf{F}\label{eq:mom}\\ 
\nabla\cdot\mathbf{u} &=& 0 \label{eq:cont},
\end{eqnarray}
Here, $\mathbf{u}(\mathbf{r},t) = [u_x(x,y,t); u_y(x,y,t)]$ is the planar velocity vector, $P(\mathbf{r},t)=p(\mathbf{r},t)/\rho_0$ is the kinetic pressure normalized by the fluid density $\rho_0$, $\nu$ is the kinematic viscosity, and $\mathbf{F}$ is an external random forcing term. Since fully developed 2D turbulence exhibits an inverse cascade, transferring energy to large scales rather than dissipating it through viscosity~\cite{Boffetta2011}, a Rayleigh-type friction term, $-\lambda\mathbf{u}$, is added to the right hand side of the momentum equation, acting as an energy sink at larger scales, mimicking unresolved dissipative physical mechanisms in the model.

Historically, truncations of NS equations have been instrumental in investigating the intricate transition to chaos~\cite{Brandstater1983,Doherty1988,Ott2002}. Despite challenges like sensitivity to mode selection and perceived universality limitations for very small mode numbers, these reduced models successfully captured various routes to chaos (e.g., bifurcations, quasi-periodicity, intermittent chaos) and reproduced experimental features~\cite{Fenstermacher1979,Gollub1980,Chen2005,Liberzon2011,Faranda2017,Carbone2020,Carbone2022b,Carbone2022c,Carbone2024}. 
Furthermore, its explicit control over the network of interacting Fourier modes and their nonlinear couplings offers direct insight into underlying dynamical mechanisms~\cite{Servidio2008A,Servidio2008B,Dmitruk2011,Gurcan2021}. 

The grid construction is a critical component, providing flexible scale resolution, effective mode decimation, and the ability to introduce a certain degree of anisotropy and ``disorder'' by specifying the ``empty'' (forbidden) zones.
Such a controlled environment, by leveraging a tunable Fourier network, 
proves particularly advantageous for studying turbulent phenomena where specific mode interactions are critical and can be precisely controlled, or even inhibited.

The primary objectives of this study are threefold: (i) to investigate the effects of anisotropy on the 2D bidirectional cascade; (ii) to examine the influence of the mode density $\rho$ (the Fourier space coverage)
in the presence of anisotropy; and (iii) to thoroughly analyze particle diffusion and pair dispersion scaling in non-uniform and disordered systems. 
%\comm{quest'ultimo pezzetto e' un pè'o confusionario. Va scritto meglio, piu enfatico, chiaro. Inoltre va messo prima nella sezione.}

\section{A $N$-mode Galerkin truncation of the two-dimensional incompressible Navier Stokes equations}
\label{sec:eq}

In wave vector space, the velocity field $\mathbf{u}(\mathbf{r},t)$ is expressed via Fourier coefficients: $\mathbf{u}(\mathbf{r},t) = \sum_k \mathbf{u}(\mathbf{k},t)e^{-i\mathbf{k}\cdot\mathbf{r}}$, where $\mathbf{k} = 2\pi\mathbf{n}/L$ with $\mathbf{n}\in\mathbb{N}$ a pair of integers. Due to the incompressibility of the fluid, these coefficients can be defined by a complex amplitude $u_\mathbf{k}(t)$ and a unit polarization vector $\mathbf{e}(\mathbf{k})$ perpendicular to the wave vector (i.e., $\mathbf{k}\cdot\mathbf{e}(\mathbf{k})=0$). Thus, $\mathbf{u}(\mathbf{k},t) = {u}_\mathbf{k}(t)\mathbf{e}(\mathbf{k})$. The unit vector $\mathbf{e}(\mathbf{k})$ satisfies $\mathbf{e}(\mathbf{k})=\mathbf{e}^\star(-\mathbf{k})$ and $\mathbf{e}(\mathbf{k})\cdot\mathbf{e}^\star(\mathbf{k})=1$, and is defined as $\mathbf{e}(\mathbf{k}) = i k^{-1}(k_y,-k_x)$, being $k_x$ and $k_y$ the components of $\mathbf{k}$ on the plane, and $k= |\bf{k}|$.

Projecting the NS equations onto Fourier space yields an infinite set of ordinary differential equations for the complex amplitudes $u_\mathbf{k}(t)$:

\begin{eqnarray}
\frac{d{u}_\mathbf{k}(t)}{d t} &=& \mathcal{N}(t) - \left[\nu k^2 + \lambda\right] u_\mathbf{k}(t) + F(k), \label{eq:ODEu} \\
\mathcal{N}(t) &=& \frac{4\pi^2}{L^2} \;\sum_{\mathbf{p},\mathbf{q}}\delta_{\mathbf{k},\mathbf{p}+\mathbf{q}}\;
C_{\mathbf{k}\mathbf{p}\mathbf{q}}[u_\mathbf{p}(t)\;u_\mathbf{q}(t)], \label{eq:ODEuZ}
\end{eqnarray}
with the conditions $u_{-\mathbf{k}}(t) = u^{\star}_\mathbf{k}(t)$.

{
The quadratic nonlinearity $\mathcal{N}(t)$ couples each mode $\mathbf{k}$ to all wave vector pairs $(\mathbf{p},\mathbf{q})$ satisfying the condition $\mathbf{k}=\mathbf{p}+\mathbf{q}$. 
As a consequence, the sum over $\mathbf{p}$ and $\mathbf{q}$ formally extends over all wave vector triads satisfying this triangular condition.
The strength of the nonlinear interaction is quantified by the coupling coefficients 
$C_{\mathbf{k}\mathbf{p}\mathbf{q}}=\tfrac{1}{2}\!\left(M_{\mathbf{k}\mathbf{p}\mathbf{q}}+M_{\mathbf{k}\mathbf{q}\mathbf{p}}\right)$, where 
$M_{\mathbf{k}\mathbf{p}\mathbf{q}}=\big[-i\,\mathbf{k}\!\cdot\!\mathbf{e}(\mathbf{q})\big]\big[\mathbf{e}^\star(\mathbf{p})\!\cdot\!\mathbf{e}(\mathbf{q})\big]$.

Here, however, the system of equations~(\ref{eq:ODEu})--(\ref{eq:ODEuZ}) is restricted to a large but finite number of modes $N_k$, yielding a finite-dimensional Galerkin-truncated system of ordinary differential equations for the retained Fourier coefficients.
Only interactions among the retained Fourier modes are included, since wave vectors outside the selected set do not appear in the dynamics. 
Accordingly, the summation in $\mathcal{N}(t)$ is effectively restricted to triads formed exclusively by retained modes satisfying the triangular condition. 
As a consequence, no new Fourier modes can be populated during the time evolution, and no discarding or filtering of modes is performed at any time step.
}

Furthermore, the number of modes $N_k$ is sufficiently large to ensure the ergodic convergence of the dynamics, thereby facilitating a robust characterization of the energy cascade. The inclusion of additional modes does not modify the overall dynamics, but rather results in a smoother Fourier spectrum in their temporal averages.

The external forcing term, $F(k)$, is a white-in-time delta-correlated random process with a prescribed power spectral density (PSD) $\mathcal{P}(k)$, and random phases $\phi_k\in[0,2\pi]$. It acts within a narrow range of wave vectors by injecting energy into the system at a rate $\epsilon_f$: 
\begin{eqnarray}
\mathcal{P}(k) &=& \frac{A_0}{f_0}\exp\left[-\left(\frac{k}{k_\textnormal{low}}\right)^{-8}-\left(\frac{k_\textnormal{high}}{k}\right)^{8}\right], \label{eq:Fk_psd}\\
F(k) &=& \sqrt{\mathcal{P}(k)}\exp\left[i\phi_k\right]. \label{eq:Fk}
\end{eqnarray}
The forcing is symmetric around the scale $k_c = (k_{\text{low}} + k_{\text{high}})/2$, $A_0$ is the unit area PSD normalization, and $f_0$ is the forcing amplitude, varied to ensure comparable average kinetic energy $\epsilon_k(t)$ (Figure~\ref{fig:energy_anisotropy} upper panel).

Since the characteristic viscous dissipation scales and friction dissipation scales can be expressed as functions of the viscous dissipation rate of energy (or enstrophy) and the friction dissipation rate~\cite{Boffetta2010}, specifically tuning the energy injection rate $\epsilon_f$ while keeping $\lambda$ and $\nu$ constant necessarily implies fixing the dissipation rates value. Consequently, this also fixes the system's characteristic dissipation scales, rendering them, to some extent, 
independent of the total number of modes, $N_k$, in the system. 

\section{Active triads and random wave-vector selection}

{
The selection of wave vectors entering the Galerkin-truncated dynamical system is performed in Fourier space within a bounded domain. Specifically, all retained modes satisfy the relation:
\[
|\mathbf{k}| \le \mathcal{K}_{\textnormal{m}},
\]
{where $\mathcal{K}_{\textnormal{m}}$ defines the maximum resolved wave vector of the truncated system, thereby setting the overall spectral extent of the dynamics and representing the smallest spatial scales explicitly retained in the Galerkin truncation. The value of $\mathcal{K}_{\textnormal{m}}$ is chosen sufficiently large to ensure that enstrophy is transferred toward the smallest resolved scales without artificial accumulation due to spectral truncation.}
}

{
Within this domain, the $(k_x,k_y)$ plane is partitioned into $N_S=14$ spectral bands, and each band sample a specific range of wave vectors. 
}

{
Each band, labeled with index $i$, is characterized by two parameters: a starting wave vectors scale $k^i_{\textnormal{start}}$ and an ending wave vectors scale $k^i_{\textnormal{end}}$.
The starting value of the first ($i=1$) band is fixed as $k^1_{\textnormal{start}}=0$. For all subsequent bands, the starting wave vectors is defined recursively as
\[
k^i_{\textnormal{start}} = \left\lceil 0.75\, k^{i-1}_{\textnormal{end}} \right\rceil ,
\]
so that consecutive bands partially overlap in wave vectors space. This overlap is introduced deliberately in order to avoid sharp gaps between scales and to ensure continuity of spectral coverage.
}

{
Within each band, wave vectors are selected through a random sampling procedure subject to an additional spacing constraint. Specifically, each band $i$ is associated with a spacing parameter $d_i$, which prescribes the minimum allowed distance between any two selected wave vectors in the $(k_x,k_y)$ plane. In other words, this constraint effectively defines the ``inter-band'' sampling resolution and prevents excessive clustering of modes. The values of $k^i_{\textnormal{end}}$ and $d_i$ for all bands are fixed a priori and are reported in Table~\ref{tab:distances}.
}

{Here and in the following, the term ``sampling'' is used in the statistical sense of randomly selecting wave vectors from Fourier space in order to define the reduced Galerkin basis. It should not be confused with the sampling of a continuous physical-space field on a grid, nor with the reconstruction of such a field from discrete spatial samples.}

{
Because different bands overlap, the same wave vector may be selected more than once during the sampling process. 
In the final construction of the truncated system, only unique wave vectors are retained: duplicated entries corresponding to identical $(k_x,k_y)$ components are discarded, so that each wave vector appears only once in the final set of retained modes. 
Nevertheless, the presence of overlap regions facilitates continuity across band interfaces and helps maintain dynamical connectivity among the retained modes. 
These connected regions, containing wave vectors common to adjacent bands, allow energy and enstrophy injected by the external forcing term $F({k})$ to be redistributed throughout the truncated system, preventing the formation of isolated or dynamically inactive regions in spectral space. 
Finally, the zero mode $(0,0)$ is always excluded by construction.
}

{
As an illustrative example, consider bands $i=8$ and $i=9$. Band $i=8$ spans wave vectors values from $k^8_{\textnormal{start}}=75$ to $k^8_{\textnormal{end}}=200$ and employs a minimum spacing $d_8=5$, resulting in a relatively coarse sampling at those scales. The subsequent band starts at:
\[
k^9_{\textnormal{start}} = \left\lceil 0.75 \times 200 \right\rceil = 150,
\]
and extends up to $k^9_{\textnormal{end}} = 300$, with a different spacing $d_9$. By combining multiple overlapping bands with scale-dependent resolutions, the procedure achieves a broad spectral coverage ranging from the largest to the smallest resolved wave vector while maintaining a controlled number of retained modes.
}

\begin{figure}[ht]
	\includegraphics[scale=0.28]{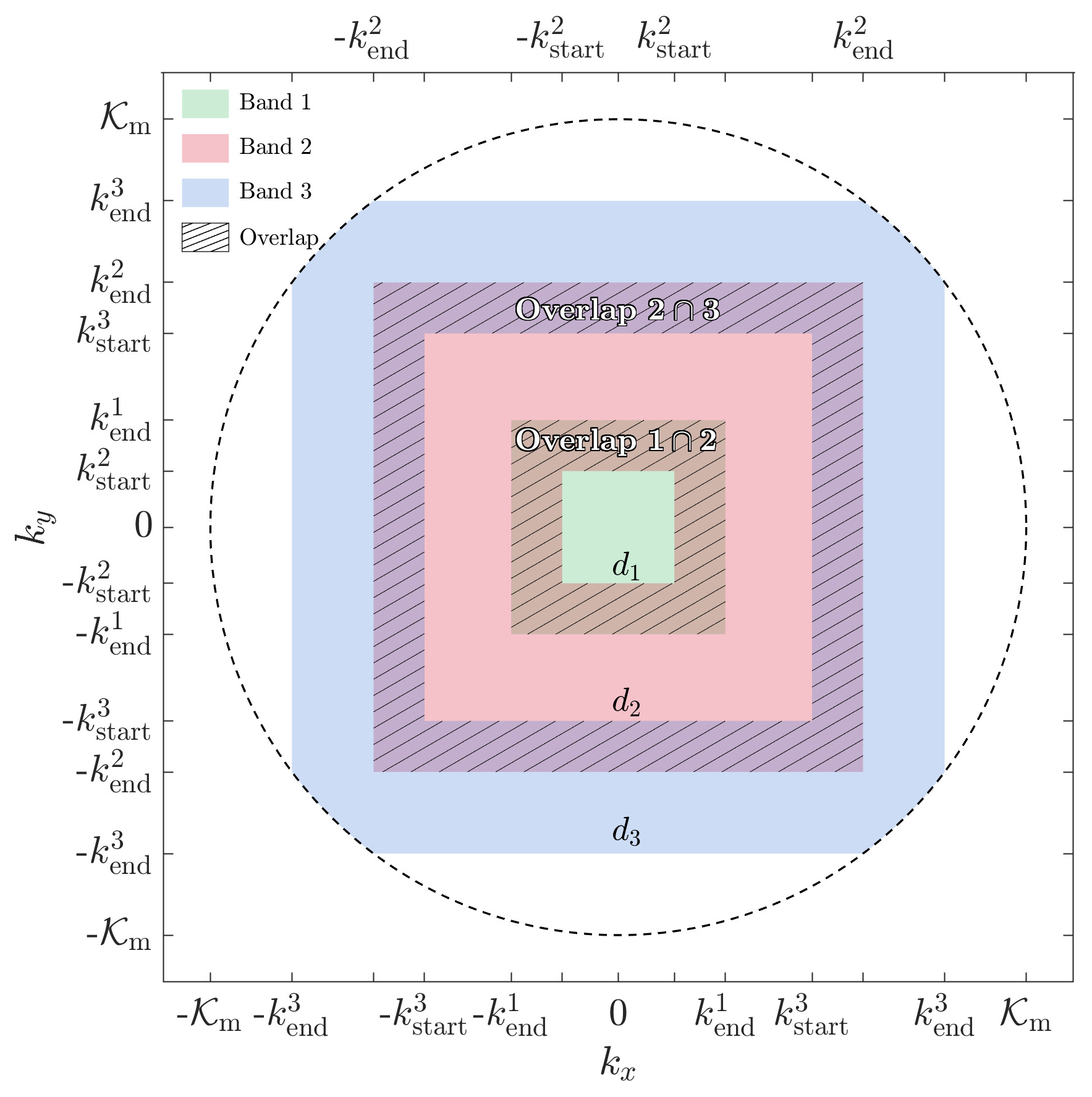}
	\caption{
		Schematic representation of the subdivision of the $(k_x,k_y)$ Fourier plane into three color-coded spectral bands within the circular truncation $|\mathbf{k}|\le \mathcal{K}_\textnormal{m}$. 
		The hatched regions indicate the overlap domains $1\cap2$ and $2\cap3$, where modes belong simultaneously to two adjacent bands. 
		Within each band, Fourier modes are sampled on a Cartesian lattice with characteristic (uniform) spacing $d_i$, with increasing coarseness across bands, $d_1 < d_2 < d_3$.
	}
	\label{fig:sketch} 
\end{figure}

{
A schematic illustration of the band-based division procedure in the $(k_x,k_y)$ plane is provided in Figure~\ref{fig:sketch}, where the global domain and the spectral bands are shown, and adjacent bands are allowed to overlap.  
In the this sketch, the Fourier plane is partitioned into three spectral bands, each characterized by an internal sampling spacing $d_i$. 
}

{
The spacing increases from the innermost to the outermost band, reflecting a progressive coarsening of the sampling as bands cover regions associated with larger wave-number magnitudes, i.e.\ $d_1 < d_2 < d_3$. 
}

\begin{figure*}[t]
	\includegraphics[scale=0.41]{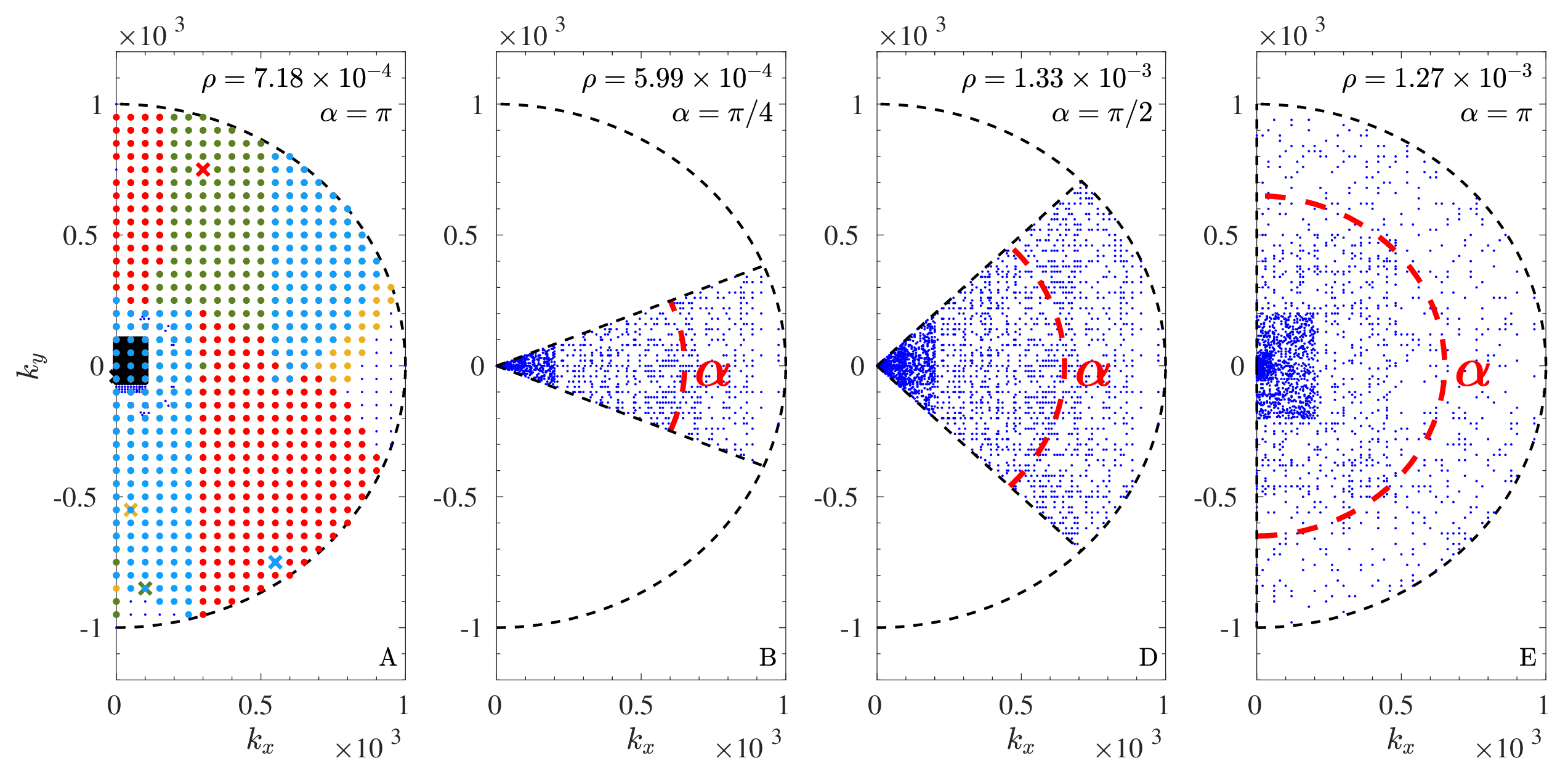}
	\caption{
		The distribution of selected wave vectors in different cases is as follows:
		{First panel:} Case $A$, the Fourier space is sampled with an ordered pseudo-logarithmic spacing, where, $k^i_{\text{end}}$ is set to $10$, $100$, and $1000$, and $d_i$ to $1$, $10$, and $50$, respectively. 
		This represents the isotropic case with lower density. {Colors indicate the sets of Fourier modes contributing to the nonlinear evolution of selected wave vectors (marked by the symbol $X$), highlighting the underlying network of triadic interactions.}
		{Second and third panels:} Two cases with different anisotropy values: case $B$ with $\alpha = \pi/4$, representing the case with higher anisotropy, and case $D$ where $\alpha = \pi/2$, representing the case with lower anisotropy.
		{Fourth panel:} Isotropic case with $\alpha = \pi$, with random spacing as reported in table~\ref{tab:distances}. This represents the highest density case, $\rho$.
		The wave vector density values are reported in Table~\ref{tab:sims}.		
	}
	\label{fig:cones}
\end{figure*}  

{
The hatched regions in Figure~\ref{fig:sketch} denote the overlap regions of the plane: $1\cap2$ and $2\cap3$, where wave vectors belong simultaneously to two neighboring bands.
}

{
Figure~\ref{fig:sketch} is intended solely as a schematic representation of the division/sampling strategy. 
The relative widths of the bands, the extent of the overlap regions, and the values of the spacings $d_i$ are not drawn to scale and do not correspond to the actual parameters employed in the simulations of the dynamical system~(\ref{eq:ODEu}). 
The purpose of the figure is to illustrate the logic of the band-based selection procedure rather than the specific numerical values used in the computations. The real values of $k^i_{\textnormal{start}}$, $k^i_{\textnormal{end}}$ and $d_i$ used in the computations are reported Table~\ref{tab:distances}.
}

In addition to these randomly sampled cases, an ordered (isotropic) pseudo-logarithmic lattice spacing case was also studied. In this scenario, $k^i_\textnormal{max}$ was set to $10, 100, 1000$ and $d_i$ to $1, 10, 50$ respectively. 
These specific values, which are not listed in Table~\ref{tab:distances}, correspond to the parameters used for the run referred to as Case $A$ in the text.

\begin{table}
	\caption{\label{tab:distances}%
		starting scale $k^i_\textnormal{start}$, ending scale $k^i_\textnormal{end}$, and internal spacing $d_i$, for each band used for the numerical integration of the system~(\ref{eq:ODEu}). Symbol $\lceil\cdot\rceil$ used in the text represents the ``ceil'' operation.
	}
	\begin{ruledtabular}
		\begin{tabular}{l|cccccccccccccc}
			band & 1 & 2 & 3 & 4 & 5 & 6 & 7 & 8 & 9 & 10 & 11 & 12 & 13 & 14  \\
			\hline
			$k^i_{\textnormal{start}}$ & 0 & 3 & 4 &  6 &  8 & 23 &  38 &  75 & 150 & 225 & 300 & 375 & 525 &  675 \\
			$k^i_\textnormal{end}$     & 3 & 5 & 7 & 10 & 30 & 50 & 100 & 200 & 300 & 400 & 500 & 700 & 900 & 1000 \\
			$d_i$ &1 & 1 & 1 & 2 & 2 & 4 & 5 & 5 & 20 & 20 & 30 & 20 & 20 & 60 \\
		\end{tabular}
	\end{ruledtabular}
\end{table} 

{
	After the identification of all admissible triads, a filtering step is applied to ensure the dynamical connectivity of the system. In particular, modes that do not participate in any triadic interaction, as well as modes involved in only a single (isolated) triad, are systematically removed. This procedure guarantees that each retained mode is embedded in a sufficiently connected network of interactions, allowing for effective spectral transfer of energy and enstrophy across scales.
	
	The total number of initially sampled modes is typically large; however, this number is reduced by the filtering step to $N_k$, which denotes the number of wave vectors that survive and therefore define the effective number of degrees of freedom of the truncated dynamical system.
}

{
The purpose of dividing the spectral domain into bands is not to impose any restriction on triadic interactions, which remain fully determined by the geometric closure condition, but rather to enable a controlled and scale-dependent sampling of Fourier modes.
For instance, wave vectors sampled in the outermost band may be dynamically connected, and thus form triadic interactions, with wave vectors belonging to inner bands, including the first or intermediate ones, provided that the geometric closure condition is satisfied.

More generally, many different spectral selection strategies could in principle be adopted, and the specific partition into bands does not represent a unique construction. However, in the limit in which the number of retained modes increases and the sampling becomes increasingly dense in wave vectors space, the truncated system formally approaches the full spectral formulation. In this sense, different admissible selection procedures are expected to recover the full spectral dynamics in the limit of dense spectral sampling, namely as $N_k \to \infty$.
}
  
The mode density, $\rho$, quantifies the sampling coverage. It's defined as the ratio of the total number of sampled wave vectors ($N_k$) to the maximum possible number of modes ($N_{\text{total}}$) that could be sampled with a lattice separation $d_i = 1$ in every band: $\rho = N_k/N_{\text{total}}$. The sampling density can also be controlled by prescribing the number of sampled points at a specific radius $k$; here, this number is set to $1$. This means only one mode is selected for each radius $k$, maintaining a very low density state (table~\ref{tab:sims}). 

In real physical system, anisotropy often arises from a change in the characteristic timescales of the flow as a function of direction. For example, large-scale shear or the presence of a strong background magnetic field can modify the turbulent cascade, causing it to be faster or slower in specific directions.
Here, anisotropy can be introduced into the turbulent dynamics by sampling only specific angular sectors (or ``slices'') of the $(k_x,k_y)$ plane within each band.
By effectively suppressing interactions in specific angular regions of the plane, the energy cascade is altered, flowing in different directions at different rates.  
These ``empty'' regions are represented by the white areas within the circular sampling domain shown in Figure~\ref{fig:cones}. For this purpose, an angle $\alpha \in [\pi/4, \pi]$ defines the width of the circular sector under consideration.

The negative portion of the plane is not reported in Figure~\ref{fig:cones} because it represents the complex conjugates portion of the plane.

{After sampling the wave vector space, the geometric resonance condition $\mathbf{k}=\mathbf{p}+\mathbf{q}$ is verified using a ``brute-force'' approach, where every possible combination of three wave vectors is tested to satisfy this criterion. Although this method is computationally expensive, the cost is associated only with the preprocessing stage in which the active triads are identified, and not with the subsequent time integration of the reduced dynamical system. Its exhaustive nature guarantees that every valid combination is found, making the verification process complete and accurate.}

To improve computational efficiency, the verification is conducted in parallel.
The full ensemble of wave vectors $\mathbf{k}$ is copied into a temporary memory space $\mathbf{k}_\textnormal{copy}$ that is accessible 
to all threads. The subsequent workload is divided among a 
number of threads $N_T$, with each thread assigned a specific portion $\mathbf{k_T} \subset \mathbf{k}$ of wave vectors.
Each thread independently searches for all possible {active triads} by comparing its $\mathbf{k_T}$ portion with all pairs $\mathbf{p},\mathbf{q} \in \mathbf{k}_\textnormal{copy}$. 
This parallel search reduces the wall-clock time required for the calculation, as the complexity scales
as $\mathcal{O}(N_{\textnormal{DOF}}^2 / N_T)$.
This approach ensures that all valid interactions are captured without the performance bottleneck of a sequential calculation.
The resulting {active triads} are then saved for subsequent computations of the dynamical system~(\ref{eq:ODEu}), as they are crucial for describing the coefficients $C_{\mathbf{k}\mathbf{p}\mathbf{q}}$ in the nonlinear term.

Figure~\ref{fig:cones} illustrates the grids constructed with the described sampling algorithm after the brute-force discrimination, where the maximal radius was set to $\mathcal{K}_\textnormal{m} = 1000$ for each case. An overview of all simulation runs, including the values of $\alpha$, $\rho$, and $f_0$, is reported in Table~\ref{tab:sims}.

{
In particular, the first panel of Figure~\ref{fig:cones} shows case $A$, corresponding to the pseudo-logarithmic spacing in the isotropic configuration ($\alpha=\pi$).

Five representative wave vectors, marked by the symbol $X$, are selected: $\mathbf{K}_{25} = (0, -40)$, $\mathbf{K}_{375} = (50, -550)$, $\mathbf{K}_{530} = (100, -850)$, $\mathbf{K}_{750} = (300, 750)$, and $\mathbf{K}_{900} = (550, -750)$. For each of these modes, all Fourier components contributing to the nonlinear term $\mathcal{N}(t)$ are displayed and grouped using different colors (matching the corresponding symbol $X$), thereby identifying the triadic interactions involved.

The figure shows that these groups are not isolated, but dynamically coupled and embedded in a connected interaction network, illustrating how even sparse sampling leads to connectivity in Fourier space.
}

\begin{table}
	\caption{\label{tab:sims}%
		Overview of the different density $\rho$, anisotropy angle $\alpha$ and forcing aplitude $f_0$ used in the various runs for the integration of system of equations~(\ref{eq:ODEu}) 
	}
	\begin{ruledtabular}
		\begin{tabular}{ccccc}
			Case & $\rho$ & $\alpha$ & $f_0$ & Grid type\\
			\colrule
			$A$ & $7.18\times 10^{-4}$ & $\pi$ & 25    & Pseudo-Logarithmic\\
			$B$ & $5.99\times 10^{-4}$ & $\pi/4$ & 110 & Random \\
			$C$ & $8.14\times 10^{-4}$ & $\pi/3$ & 150 & Random \\
			$D$ & $1.33\times 10^{-3}$ & $\pi/2$ & 90  & Random \\
			$E$ & $1.27\times 10^{-3}$ & $\pi$ & 75    & Random \\
		\end{tabular}
	\end{ruledtabular}
\end{table}

{The sparse sets of wave vectors considered in this study are not designed to provide a pointwise-complete representation of arbitrary prescribed physical-space structures, as a uniformly resolved grid would. Instead, each retained set defines a distinct reduced dynamical system, with its own dynamical properties. The suitability of these reduced representations is therefore assessed a posteriori through the turbulent statistics generated by the retained nonlinear interactions, rather than through pointwise reconstruction of prescribed fields. Accordingly, the configurations $A$--$E$ reported in Table~\ref{tab:sims}, characterized by different densities, anisotropy levels, and selection strategies of retained wave vectors, provide a sensitivity test of the statistical observables with respect to changes in the reduced spectral representation.}

\section{Numerical Results for truncated low-density turbulence}

System of equations~(\ref{eq:ODEu}) is integrated in time by means of a classical fourth order Runge-Kutta temporal advancement scheme, for $\lambda t = 5.5$ time units. 
For all runs a Reynolds number $\textnormal{Re} \equiv \nu^{-1} = 1.1\times10^5$, and a friction coefficient $\lambda = 10^{-3}$ have been used. The forcing scales were set to $k_\textnormal{low} = 70$ for cases $B,C,D,E$, and $k_\textnormal{low} = 65$ for case $A$, while a value $k_\textnormal{high} = 90$ was used for all runs.
The initial condition for each run is defined by a flat energy spectrum with mean amplitude $\mathcal{U}_0 = 5 \times 10^{-2}$ and uniformly distributed random phases $\phi_k$, i.e., $E_0(k) = \sqrt{\mathcal{U}_0} \exp[i\phi_k]$.

\begin{figure}[ht]
	\includegraphics[scale=0.32]{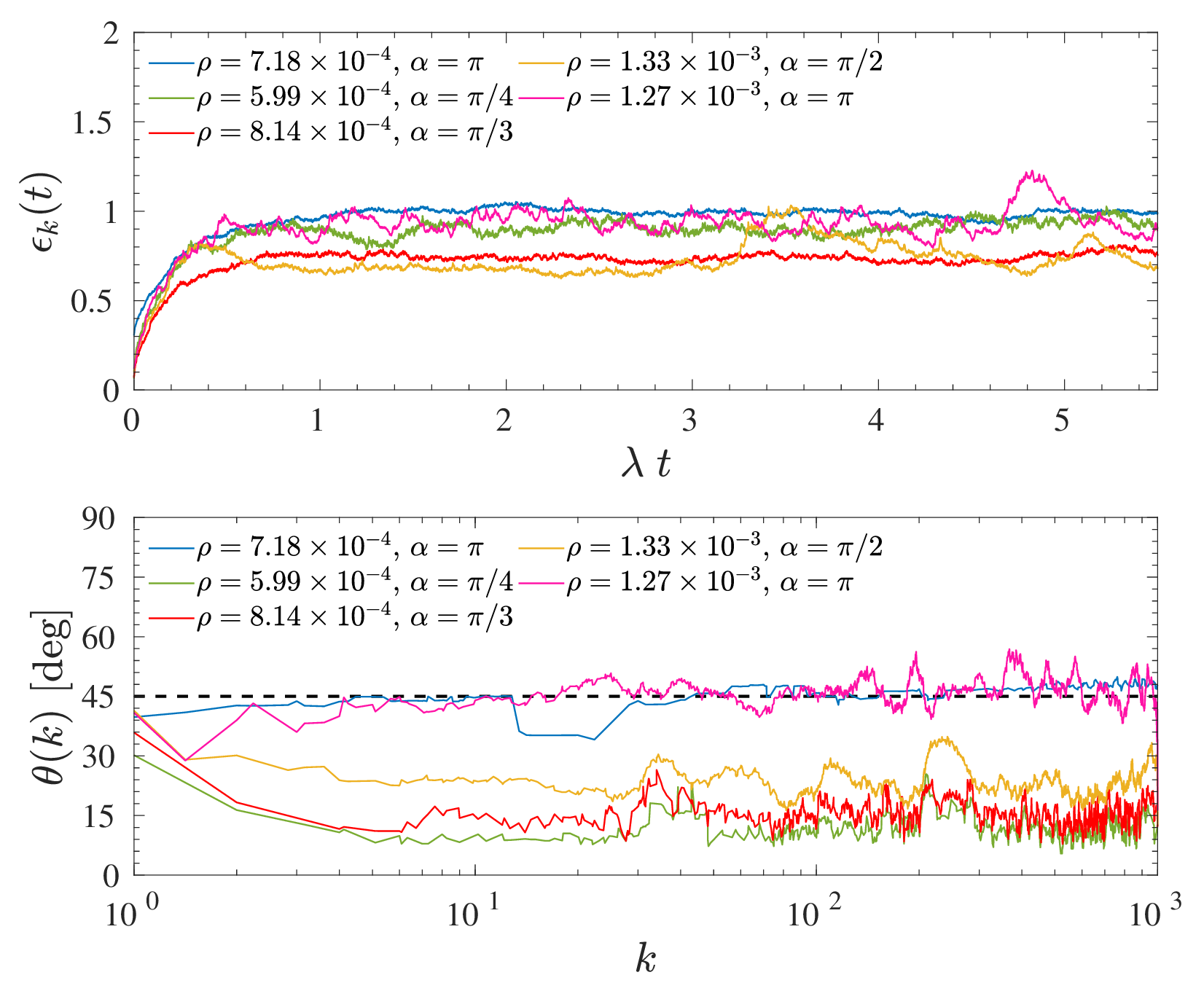}
	\caption{Upper panel: Temporal evolution of the kinetic energy $\epsilon_k(t) = \sum_k |u_\mathbf{k}(t)|^2$, for each run reported in table~\ref{tab:sims}. The amplitude of the forcing $f_0$ is tuned in each run in order to get the same value of kinetic energy.
	Lower panel: averaged spectral anisotropy angle $\theta(k)$ observed in the steady state of the dynamic. 
	For$\theta(k) = 45^\circ$ the system is isotropic (no difference in the PSD of the two component of the velocity), for $\theta(k) < 45^\circ$ the $y$ component of the field is dominant.}
	\label{fig:energy_anisotropy} 
\end{figure}

\begin{figure}[h]
	\includegraphics[scale=0.33]{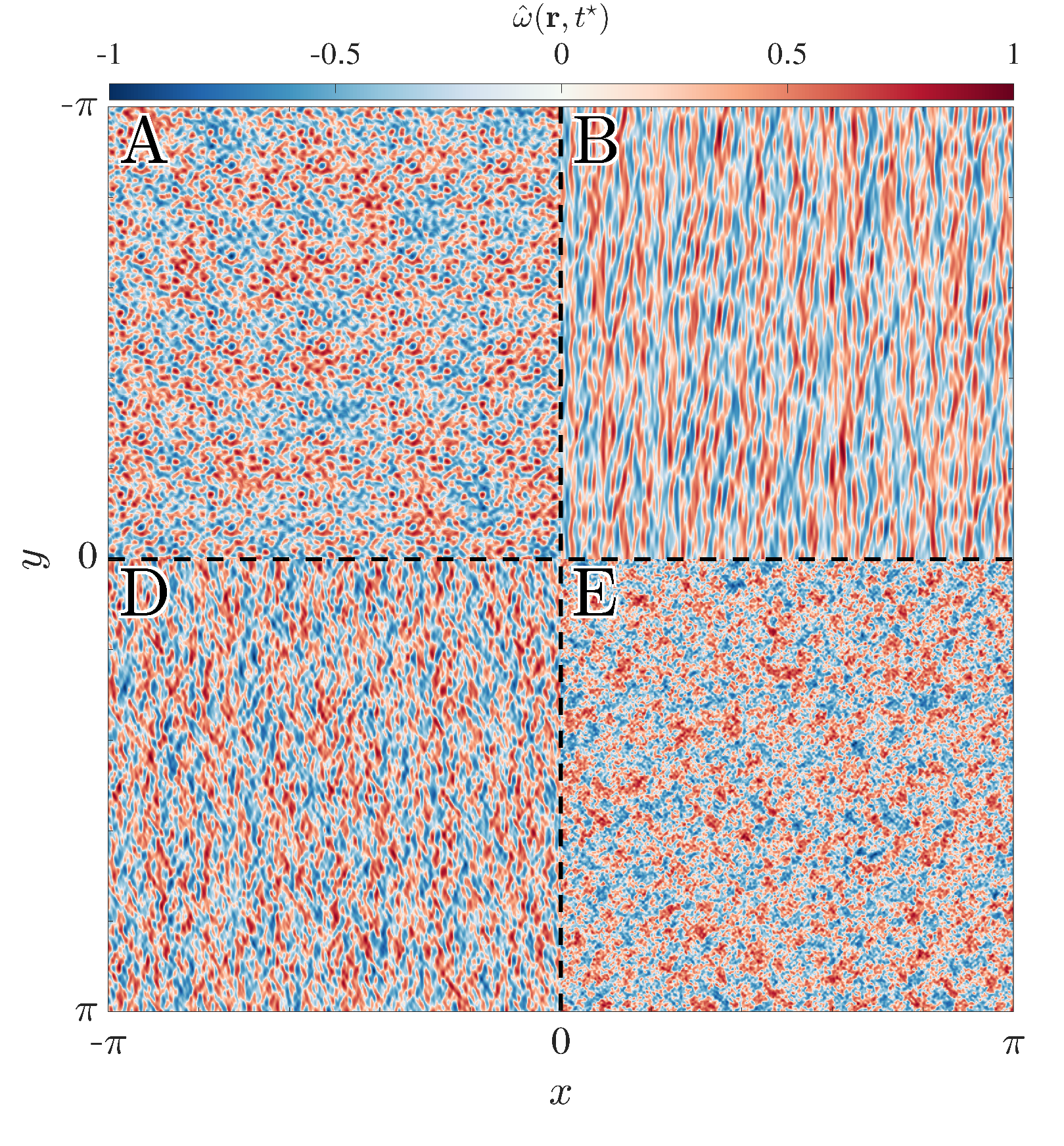}
	\caption{
		Four snapshots of the portion of the vorticity field $\omega(\mathbf{r},t^\star)$ ($x,y\in[0,\pi]$).
		The main diagonal panels present the isotropic cases $A$, and $E$, while on the secondary diagonal are reported the anisotropic cases $D$ and $B$.
		For better visualization of the field structures, each panel is normalized by $\textnormal{Max}\left\{|\omega(\mathbf{r},t^\star)|\right\}$.}
	\label{fig:vorticity}
\end{figure}

\begin{figure*}[ht]
	\centering\includegraphics[scale=0.36]{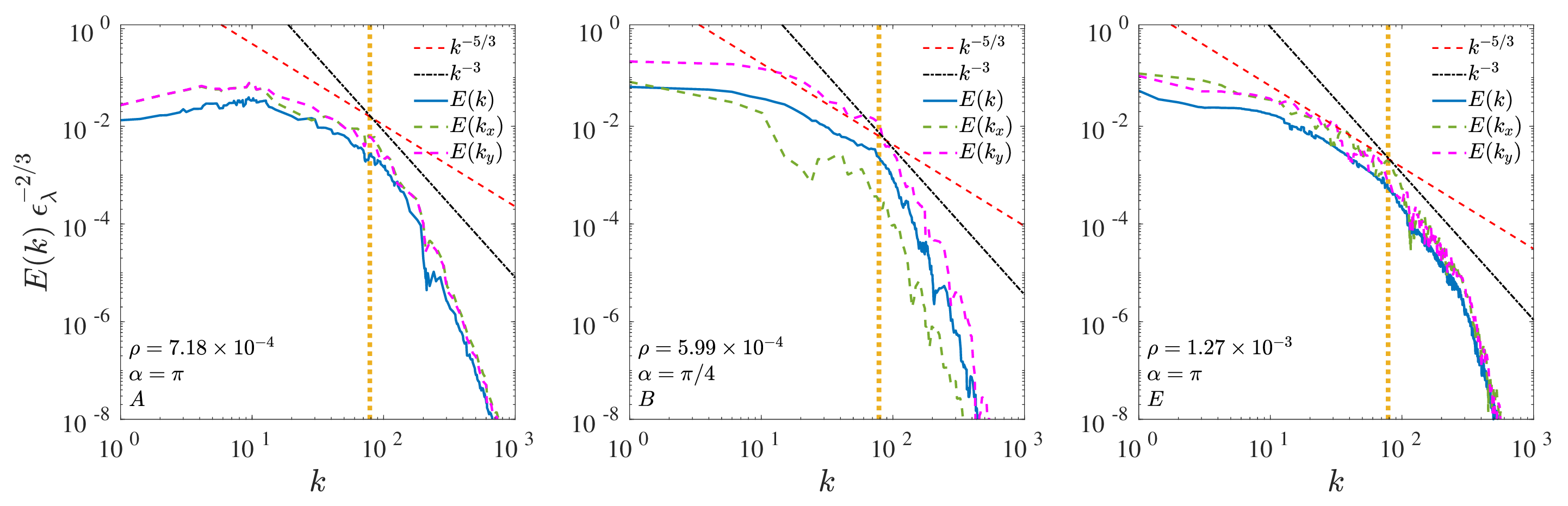}
	\caption{
		Time-averaged kinetic energy spectra $E(k)$ from three distinct runs, normalized by the friction energy dissipation rate $\epsilon_\lambda$. Left and right Panels illustrate the isotropic cases ($A$ and $E$), where $E(k_x)$ and $E(k_y)$ components show comparable power. In contrast, central panel, representing the highly anisotropic case $B$ where $\alpha=\pi/4$. This case presents a significant power discrepancy, with the $E(k_x)$ component notably lower. 
		All spectra clearly show an inverse cascade, $E^{\textnormal{inv}}(k) \sim k^{-5/3}$ (red dashed line), on the left of the forcing scale ($k<k_c$) $k_c$ (vertical dashed line), while a steeper direct cascade, $E^{\textnormal{dir}}(k)\sim k^{-3}$ (dot-dashed line), appears for $k>k_c$. 
		%The extent of the direct cascade range diminishes with increased anisotropy and strengthens as anisotropy is reduced.
		%\comm{io cambierei il rapporto di aspetto nelle figure, riducento un po l'altezza (cosi si stretchano e i range inerziali sono piu chiari). Sempre per lo stesso motivo, taglierei il kmax a 700 invece di 100.}
	}
	\label{fig:psd}
\end{figure*}

Under the effect of the forcing $F(k)$, the system reaches a statistically steady state in a relatively short time (Figure~\ref{fig:energy_anisotropy}, top panel). This constant kinetic energy ($\epsilon_k(t)$) level is maintained throughout the entire run, for all cases, up to $\lambda\,t = 5.5$. All time-averaged quantities reported in the text refer to the interval $2 \leq \lambda\,t \leq 5$. The strength of anisotropy can be quantified in terms of an angle $\theta(k)$ defined as~\cite{Shebalin1983}:
\begin{equation}
\theta(k) = \frac{180}{\pi} \tan^{-1}
\left\{ \frac{ \sum_k|\left[ \hat{\mathbf{e}}_x \cdot \mathbf{e}(\mathbf{k}) \right] u_{\mathbf{k}}|^2 }{ \sum_k|\left[ \hat{\mathbf{e}}_y \cdot \mathbf{e}(\mathbf{k}) \right] u_{\mathbf{k}}|^2 } \right\}^{1/2},
\label{eq:anis_angle}
\end{equation}
where the sum is extended over all wave vectors $\mathbf{k}$ such that $|\mathbf{k}| = k$, $\hat{\mathbf{e}}_x = (1,0)$ and $\hat{\mathbf{e}}_y = (0,1)$,
and is reported in the lower panel of Figure~\ref{fig:energy_anisotropy}.
For a purely isotropic field, $\theta(k) = 45^\circ$, as observed in the results for cases $A$ and $E$. On the other hand, the strength of anisotropy is measured by the deviation of the ratio from this value. When $\theta > 45^\circ$, the velocity component along $\hat{\mathbf{e}}_x$ dominates, whereas for $\theta < 45^\circ$, the component along $\hat{\mathbf{e}}_y$ dominates. Due to the sampling of $k_x, k_y$ and the polarization directions $\mathbf{e}(\mathbf{k})$, cases $B$, $C$, and $D$ consistently exhibit dominance of the $\hat{\mathbf{e}}_y$ component, with average values in the range $15^\circ \leq \theta(k) \leq 25^\circ$. As $k \to 0$, the system naturally tends toward isotropy, and accordingly, $\theta(k)$ approaches $45^\circ$, indicating a recovery of isotropy at large scales.

The anisotropy can be clearly quantified by observing the vorticity fields ${\omega}(\mathbf{r},t^\star)$ in physical space.
Figure~\ref{fig:vorticity} depicts four representative examples of the vorticity field $\hat{\omega}(\mathbf{r},t^\star)\equiv {\omega}(\mathbf{r},t^\star)/\omega_\textnormal{max}$ (for clarity scaled on $\omega_\textnormal{max} = \textnormal{Max}\{|{\omega}(\mathbf{r},t^\star)|\}$) obtained from the numerical integration of system~(\ref{eq:ODEu}) for $t^\star=5500$,  
reconstructed on a doubly periodic torus of length $L=2\pi$, with $N_{\textnormal{DOF}}=2048^2$.
{The vorticity fields shown in Figure~\ref{fig:vorticity} are obtained a posteriori by evaluating the finite Fourier series associated with the retained Galerkin modes. The physical-space grid is used only for visualization and does not enter the time integration of the Galerkin systems defined by Equations~\ref{eq:ODEu}, and~\ref{eq:ODEuZ}. Since the retained modes satisfy $|\mathbf{k}|\leq \mathcal{K}_{\textnormal{m}}=1000$, their Cartesian components satisfy $|k_{x}|,|k_{y}|\leq1000$ and are therefore below the Nyquist wavenumber $k_{\rm Ny}=1024$ of the reconstruction grid. Therefore, the diagnostic reconstruction resolves all retained modes and does not introduce aliasing in the reconstructed physical-space fields. Moreover, no nonlinear products are evaluated on this grid, so no spectral aliasing is introduced by the reconstruction.}
For clarity each panel shows only a portion of the vorticity field $x,y\in[0,\pi]$.
Arranged from top-left, panels along the main diagonal display the isotropic cases ($A$ and $E$), while the anti-diagonal panels correspond to two anisotropic cases ($B$ and $D$).

While both run $A$ and $E$ represent isotropic cases, they exhibit subtle yet significant differences in the structure of the vorticity field. Pseudo-logarithmic case $A$ reveals a vorticity distribution where large-scale structures (broader, coherent regions of positive or negative vorticity) are clearly distinguishable, appearing to emerge from an aggregation of smaller structures or as composite entities. Overall, the field exhibits a higher degree of spatial organization at intermediate and large scales compared to the random isotropic case $E$. Although the vortices maintain an overall isotropic distribution, they are not entirely randomly dispersed; instead, they tend to form larger, visible aggregates, delineating a distinct large-scale pattern~\cite{Kraichnan1989,Sanada1992,Carbone2011}. This morphological characteristic might indicate a system phase where the energy condensation towards large scales is favored, leading to the formation of extended coherent structures and then dissipated via Rayleigh friction terms.

For random case $E$, the vorticity field appears finely fragmented and homogeneous across all resolved scales, exhibiting a uniformly distributed and finely structured turbulence. Indeed, the vortex distribution is denser and more uniformly dispersed, showing less evidence of the large-scale ``composite'' structures 
as the pseudo-logarithmic case $A$. The overall field suggests a more homogeneous and almost isotropic turbulence state with fewer emerging coherent structures.

The effect of anisotropy is significantly more pronounced in cases $B$ and $D$. Indeed, anisotropy manifests as a distinct and preferential orientation of the vorticity structures. In contrast to the isotropic cases, elongated structures are observed along a privileged direction, thus indicating an anisotropic energy transfer. These structures often appear as alternating filamentary patterns, or bands, of positive and negative vorticity. 

In the classical double-cascade scenario~\cite{Boffetta2010}, with an extended inertial range, 
the energy injected at the forcing scale $k_c$ is rapidly transferred to large scales, where is dissipated by frictional forces, forming the inverse cascade, while all the enstrophy is transferred to small scales in the direct cascade, and then dissipated by small scale viscosity.

The presence of the {double cascade} is easily evinced by observing the system's kinetic energy spectra $E(k)$. In Figure~\ref{fig:psd} are shown the temporal averaged power spectra (PSD), resulting from three different runs, compensated with the energy friction dissipation rate $\epsilon_\lambda$. Each panel of Figure~\ref{fig:psd} reports the spectra for both velocity components $E(k_x)$, $E(k_y)$, and the isotropic spectrum, averaged on the modulus $k$, i.e., $E(k)$. The first and third panels of Figure~\ref{fig:psd} present the two isotropic cases: the pseudo-logarithmic case A and the random sampled case E, and in both, the spectra of the $x$ and $y$ components exhibit the same power. On the other hand, the central panel of Figure~\ref{fig:psd} reports the case with the strongest anisotropy $\alpha = \pi/4$; in this case, a strong discrepancy between the two components is observed, as the system's $x$ component is significantly lower in power, indicating that energy transfer along this component is disfavored compared to the $y$ direction.

\begin{figure}[ht]
	\centering\includegraphics[scale=0.45]{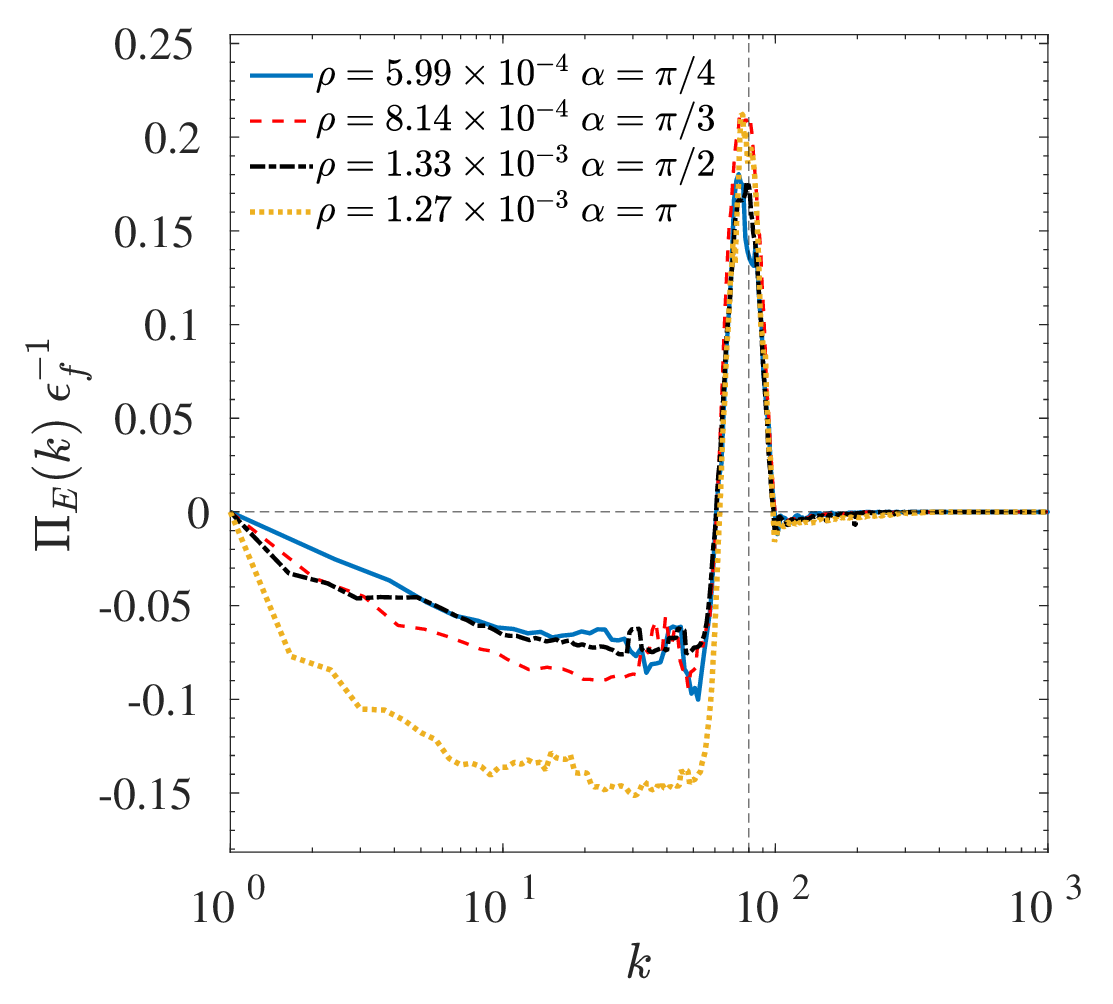}
	\caption{
		Averaged energy flow among scales, $\Pi_E(k)$, normalized by the injected energy $\epsilon_f$. All runs present an inverse energy transfer (negative values) suppressed by large-scale friction as $k \to 0$. The efficiency of the process increases as anisotropy decreases (e.g., run $E$). A constant negative energy flux plateau is visible due to the dual energy sinks (viscosity and friction). Anisotropic cases show slightly reduced inverse flow efficiency due to sub-optimal wave vector coupling; all runs consistently transfer energy from the injection scale $k_c$ (vertical dashed line) to larger scales.
	}
	\label{fig:flowpi} 
\end{figure}

\begin{figure*}[ht]
	\centering\includegraphics[scale=0.43]{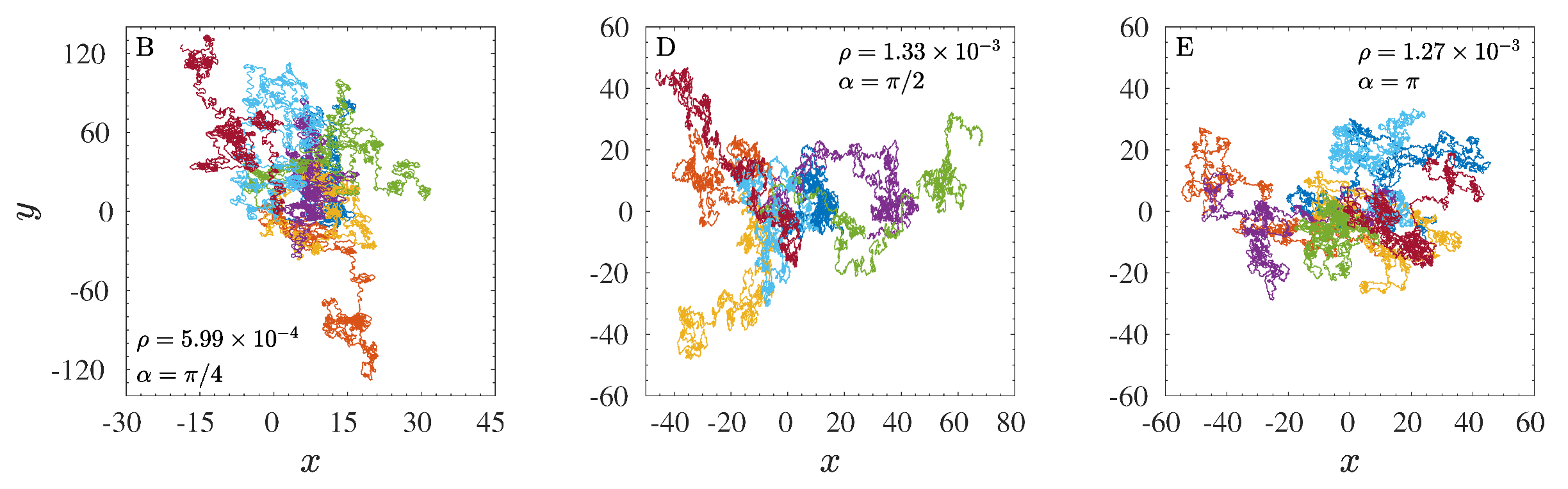}
	\caption{
Comparison of single particle trajectory for the first $5$ particle injected in the system (color coded), during three different runs case $B$ left panel, case $D$ central, and case $E$ right panel, for $1500 \leq t \leq 5500$.  
Strongly anisotropic case (left and central panel, $\alpha=\pi/4$, and $\alpha=\pi/2$) presents significantly enhanced diffusivity along the $y$-axis, differing by nearly a factor $4$ compared to the $x$-direction. This contrasts with isotropic cases (right panel, $\alpha=\pi$), where diffusivity is comparable across both directions.
	}
	\label{fig:TRAJ} 
\end{figure*}

However, despite the different power, in all cases, the classical double scaling is observed. As can be seen, to the left of the vertical dashed line, which corresponds to the central forcing scale $k_c$, a power-scaling compatible with $E^\textnormal{inv}(k) \sim k^{-5/3}$ (red dashed line) is observed, as expected from classical theory~\cite{kraichnan1965,Boffetta2010}. To the right of $k_c$, towards smaller scales, the system exhibits a steeper scaling, also compatible with classical $E^\textnormal{dir}(k)\sim k^{-3}$ (black dot-dashed line). These two ranges are observed in all cases, both isotropic and anisotropic grids. However, the portion of the spectrum affected by the direct cascade is smaller in anisotropic cases, and the size is related to the value of the anisotropy: increasing $\alpha$ (thus reducing the anisotropy) reinforces the evidence of the direct cascade. 
Although the spectral exponent in the first panel of the figure suggests that energy is transferred upscale,
the absence of a condensate implies that this energy transfer must stop before reaching $k=1$.
Such behavior is likely attributable to inefficient energy transfer, probably due to the construction of the grid in Fourier space.
The flux of energy across scale $k$ in the statistically steady state can be defined in Fourier space~\cite{Alexakis2005,Boffetta2011,Linkmann2019} as:
\begin{equation}
\Pi_E(k) = - \left\langle \int_{\mathbf{|k^\prime|\leq k}} d\mathbf{k}^\prime \tilde{\mathbf{u}}(-\mathbf{k}^\prime)\cdot\mathcal{F}_{\mathbf{k}^\prime}\left[(\mathbf{u}\cdot \nabla)\mathbf{u}\right] \right\rangle,
\label{eq:flow}
\end{equation}
where $\mathcal{F}[\cdot]$ imply Fourier transform operation.
Such relation can be easily evaluated by exploiting the conjugate conditions and the nonlinear term $\mathcal{N}(t)$ as in equation~(\ref{eq:ODEuZ}).
The sign of $\Pi_E(k)$ is defined such that $\Pi_E(k)< 0$ corresponds to an inverse energy transfer and $\Pi_E(k) > 0$ to a direct energy transfer.

As shown in Figure~\ref{fig:flowpi}, the fluxes tend to zero as $k\to0$ indicating that the inverse energy transfer is suppressed by friction dissipative effects. The effect is evident in all cases, and the efficiency of the inverse flux increases with decreasing anisotropy, such effects is particularly evident for run $E$. Moreover, normalizing the fluxes to $\epsilon_f$ demonstrates that the energy injection is constant across all runs.
Due to the presence of the two energy sinks (small-scale viscosity and large-scale friction), a plateau in the energy flow $\Pi_E(k)$ represents a constant energy flux. This behavior is observable in all runs and becomes clear for the isotropic run $E$. In contrast, for the other anisotropic cases, the inverse flow exhibits slightly reduced efficiency, as the coupling between wave vectors is {not optimal}. 
Nevertheless, all numerical observations consistently reveal that energy is transferred from the injection scale to larger scales, where it is subsequently dissipated by friction.

Even though the energy transfer is not optimal, it is a striking observation that a reduced system, where the isotropy hypothesis does not always hold, reflects a power spectrum where energy is distributed according to the classical 2D turbulence scaling laws, and, moreover, the integral properties of the energy cascade are maintained. This is particularly noteworthy when contrasted with a full DNS, where isotropy is valid by definition.

\section{Particle pair statistics in non-uniform and disordered Fourier spaces}

As known, in the study of turbulence, the Lagrangian approach provide a unique and crucial perspective by focusing on the motion of individual fluid particles as they are swept along by the flow. While traditional Eulerian statistics are valuable, they fail to fully capture the highly intermittent and chaotic nature of turbulent transport. This is where Lagrangian analysis becomes essential, as it directly addresses how fluid elements are stretched, rotated, and diffused.
 
In classical turbulence theory, it's expected that particle pairs separate diffusively over time. However, due to the intermittency of the velocity field, many experimental and numerical studies have revealed a phenomenon where the separation of particle pairs follows an anomalous power-law (e.g. Richardson scaling~\cite{Richardson1926}), deviating significantly from the expected diffusive behavior. This anomalous separation is a direct manifestation of the non-Gaussian nature of turbulent fluctuations and provides a powerful tool for understanding how energy is transferred across scales, ultimately leading to a more complete picture of turbulent mixing and dispersion.

To investigate the diffusive properties of the truncated model, the system~(\ref{eq:ODEu}) is augmented with $N_p = 2^{14}$ equations describing the Lagrangian trajectories $\mathbf{x}_1(t), \mathbf{x}_2(t), \dots, \mathbf{x}_{N_p}(t)$ of $N_p$ passive tracer particles:
\begin{equation}
\frac{d \mathbf{x}_n(t)}{dt} = \sum_{j = 1}^{N_k} u_{j}(t) \mathbf{e}(\mathbf{k}_j) e^{i \mathbf{k}_j \cdot \mathbf{x}_n(t)},
\label{eq:dinamica}
\end{equation}
where the subscript $n$ identifies the $n$-th particle advected by the velocity field.

Particle initial positions are randomly distributed in space, with initial pair separations characterized by a parameter $R_0$, uniformly sampled in the range $R_0 \in [0, R_\textnormal{max}]$, representing the initial average separation among particles.

All particles are released at time $\lambda t = 1.5$, once the system has reached a statistically steady state, and different values of $R_0$ are considered.

{
Classical descriptions of particle pair diffusion often rely on the assumption of local isotropy. 
In systems characterized by a very low mode density $\rho$, however, this assumption may not be satisfied, and anisotropic features can persist over a wide range of scales.
}

{	
This behavior is illustrated by the trajectories of test particles in the $(x,y)$ plane shown in Figure~\ref{fig:TRAJ}. 
As the anisotropy parameter $\alpha$ is reduced (left and central panels, corresponding to cases $B$ and $D$), particle motion exhibits a markedly enhanced dispersion along the $y$ direction. 
In the strongly anisotropic case $\alpha=\pi/4$, the difference between the two directions reaches approximately a factor of four.
By contrast, in the isotropic configuration $E$ ($\alpha=\pi$), comparable diffusivities are observed along both spatial directions (Figure~\ref{fig:TRAJ}, right panel).
}

\begin{figure*}[ht]
	\includegraphics[scale=0.365]{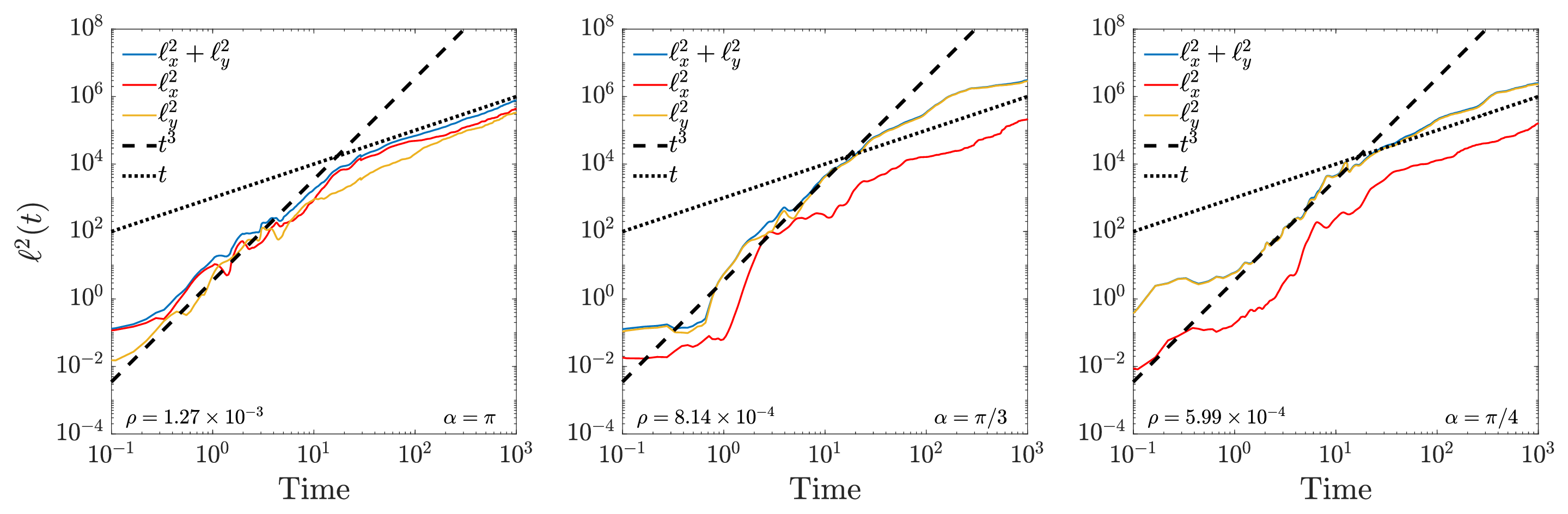}	
	
	\caption{
		Temporal evolution of the relative dispersion tensor components, averaged over all particles.
		Left panel: Isotropic case $E$ with initial separation $R_\textnormal{max} = 10^{-2}$; Central Panel: strongly anisotropic case $B$ with $R_\textnormal{max} = 10^{-2}$; Right panel: anisotropic case $D$ with $R_\textnormal{max} = 10^{-4}$. For all cases examined, the two components of the dispersion tensor reveal a dual scaling behavior: a superdiffusive regime characterized by $\ell^2 \sim t^3$ for short timescales (dashed line), transitioning to a Brownian regime defined by $\ell^2 \sim t$ for longer timescales (dotted line).
	}
	\label{fig:ell2}
\end{figure*}

{	
In addition, particle trajectories reveal frequent trapping and re-trapping events associated with coherent vortical structures.
This behavior reflects the organization of the velocity field on relatively small-scale structures, at least along one spatial direction, and provides a qualitative indication of how anisotropy influences transport properties in the present system.
}

{
The key statistical quantity used to characterize particle pair separation is the relative dispersion tensor:
$\sigma_{ij}(t) = \langle r_i(t)\, r_j(t) \rangle_{n,m}$,
where $\mathbf{r}(t) = \mathbf{x}_n(t) - \mathbf{x}_m(t)$ denotes the separation vector between a generic pair of particles at time $t$.
This tensor provides a compact statistical measure of pair dispersion and is commonly employed in both experimental and numerical studies of turbulent transport.
In classical turbulence theory, it is related to the probability density function of particle separations through a master equation framework~\cite{Richardson1926,MoninYaglom}, although in the present study it is used as a diagnostic observable to quantify dispersion properties of the truncated system.
}

{
Super-ballistic pair dispersion may also emerge as an empirical property of the dynamics,
even in reduced and anisotropic systems.
In the classical framework of isotropic turbulence, the averaged one-dimensional squared pair separation
$\ell_j^2(t)=\langle r_j(t) r_j(t)\rangle$ is identical in all directions
($\ell_x^2\simeq \ell_y^2$), and the off-diagonal components vanish.
In this case, particle pair dispersion is commonly characterized by the total squared separation
$\ell^2(t)=\ell_x^2+\ell_y^2$.
Within phenomenological descriptions, the probability density of particle separations
$p(\ell,t)$ can be modeled through a non-Fickian diffusion equation with an effective eddy diffusivity
$\eta(\ell,t)$~\cite{Bourgoin2015}.
}

{
If one further assumes that $\eta(\ell,t)$ depends only on the turbulent energy injection rate
$\epsilon_f$ and on the instantaneous separation $\ell(t)$, the well-known Richardson scaling
$\ell^2(t)\sim t^3$ follows.
This result is associated with Kolmogorov-type velocity scaling,
$\delta u_\ell\sim \ell^{1/3}$~\cite{K41}, is expected to hold within the inertial range of fully
developed turbulence, and implies the presence of a viscous anomaly~\cite{Frish1995}.
}

{
This regime is valid for timescales within the inertial range of turbulence, specifically for $t_0 \ll t \ll t_\mathcal{L}$, where $t_0$ and $t_\mathcal{L}$ are the Kolmogorov microscale and the injection scale, respectively~\cite{Taylor1938}.
}

{
For larger scales ($t > t_\mathcal{L}$), particle pairs are completely separated, behaving like independent Brownian particles without correlation. In this context, assuming $\eta$ depends on the large-scale separation eddies $\mathcal{L}$ and the average energy injection rate, a Brownian diffusive process for pairs is obtained, where $\eta = \eta_0 \mathcal{L}^{4/3} \epsilon^{1/3}$ and $\ell^2(t) = \ell_1 t$ ($\ell_1$ is a constant)~\cite{Jullien1999,Ishihara2002}.
}

{
This discussion is meant to place the numerical results reported below into a broader phenomenological context, showing that super-ballistic pair dispersion may arise as an empirical feature of the dynamics even in reduced and anisotropic systems.
}

It is worth noting that the $K41$ theory does not predict a unique diffusion coefficient~\cite{Klafter1987}. Consequently, a scaling relation for the eddy diffusivity can be introduced by retaining an explicit dependence on time $t$ and initial separation $\ell_{ij}(t=0)$, such that $\eta(\ell_{ij}, t)\simeq \eta_0\ell_{ij}^{\alpha_{ij}}(0)\epsilon^{\beta_{ij}}t^{\zeta_{ij}}$, where the scaling exponents can be determined from dimensional considerations as $\alpha_{ij} = 2-2(1+\zeta_{ij})/3$ and $\beta_{ij} = (1 + \zeta_{ij})/3$. The time scaling exponent $\zeta_{ij}$ remains, however, undetermined. Following this, the squared separation tensor can be expressed as $\ell_{ij}^2(t) = g_{ij}\ell_{ij}^{\alpha_{ij}}(0)\epsilon^{\beta_{ij}}t^{1+\zeta_{ij}}$. For the isotropic case, the Richardson regime is observed when the mean squared separation is independent of the initial separation (i.e., $\alpha_{ij} = 0$), which is obtained when $\zeta_{ij} = 2$ and $\beta_{ij}=1$.

In each run, the full tensor $\sigma_{ij}(t)$ is evaluated at each time step of the integration by measuring the one-dimensional distances $r_x(t) = x_n(t) - x_m(t)$ and $r_y(t) = y_n(t) - y_m(t)$ between $n$-th and $m$-th pairs, and then by averaging over all pairs:
$\sigma_{ij}(t) = \langle r_i(t) r_j(t) \rangle$, so that the one-dimensional components of the tensor $\sigma$ are $\ell_x^2(t) = \sigma_{xx}$ and $\ell_y^2(t) = \sigma_{yy}$, and the off-diagonal element $\ell_{x,y}(t) = \sigma_{x,y}$.

{
A close examination of the particle trajectories shows that when a particle is expelled from a vortex-like structure, it typically travels only a limited distance before being rapidly re-trapped by a nearby one, as a consequence of the spatial organization of the velocity field. As a result, particle motion is characterized by a continuous alternation of trapping and ejection events, during which particles experience frequent and substantial changes in their velocity.
}

{
This dynamics leads to an enhanced separation of particle pairs over time. In particular, the measured pair dispersion exhibits a growth consistent with a Richardson-like behavior over an extended temporal range, as illustrated in Figure~\ref{fig:ell2}. This observation is purely phenomenological and does not imply a derivation of Richardson scaling. Rather, it highlights that features commonly associated with turbulent pair dispersion can emerge even within  reduced model.
}

{
It is worth noting that particle pair dispersion has previously been investigated in low-dimensional or truncated dynamical systems~\cite{Crisanti1990,Crisanti1991,Bohr2005,Carbone2022c}, where Richardson-like scaling was not observed. 
The fact that such scaling becomes evident in the present framework is therefore nontrivial and suggests that the number and distribution of retained modes may play a crucial role in shaping particle dispersion properties. 
In particular, these results are consistent with the possible existence of a critical mode density above which super-ballistic dispersion effects become observable, even in truncated dynamical systems.
{ 
In other words, when the mode density is too low, the triadic network may fail to transfer energy or enstrophy efficiently across scales. If the network is too sparse, energy may be rapidly channelled through a few available pathways toward the lowest retained scales, preventing the formation of an extended inverse cascade. Conversely, when interacting modes have strongly different nonlinear time scales, smaller scales may act as localized reservoirs: higher $k$ may evolve on time scales that are not dynamically commensurate with those of slower large-scale modes, reducing the effectiveness of nonlinear coupling. In both cases, the cascade is weakened or partially blocked, suggesting that a sufficient density/connectivity threshold is required to sustain robust cascade dynamics and the associated super-ballistic dispersion.
}
}

The various component of the tensor $\sigma_{ij}$ (i.e. $\ell_x^2 + \ell_y^2$, $\ell^2_x$, $\ell^2_y$) are reported in Figure~\ref{fig:ell2}, for two different values of initial separation and different anisotropy values 
$\alpha$: $R_0 = 10^{-2}$ (case $E$,$B$), left and central panel, and $R_0 = 10^{-4}$ (case $D$), right panel.
For smaller values of $R_0$ (Figure~\ref{fig:ell2}, right panel), This is likely due to the weak intermittency effects observed at the present level of forcing, and the wider distribution of scales. Only after a certain time ($t \approx 0.9$, Figure~\ref{fig:ell2}, right panel) the separation becomes important, and the super-ballistic scaling is retrieved.
As expected, when the particle pair moves relatively without correlation (e.g. at longer times, $t >100$), the system also reaches a Brownian-like diffusion regime characterized by a scaling for the various components of the dispersion tensor of the form $\ell^2 \sim t$.

\begin{figure}[h]
	\centering\includegraphics[scale=0.45]{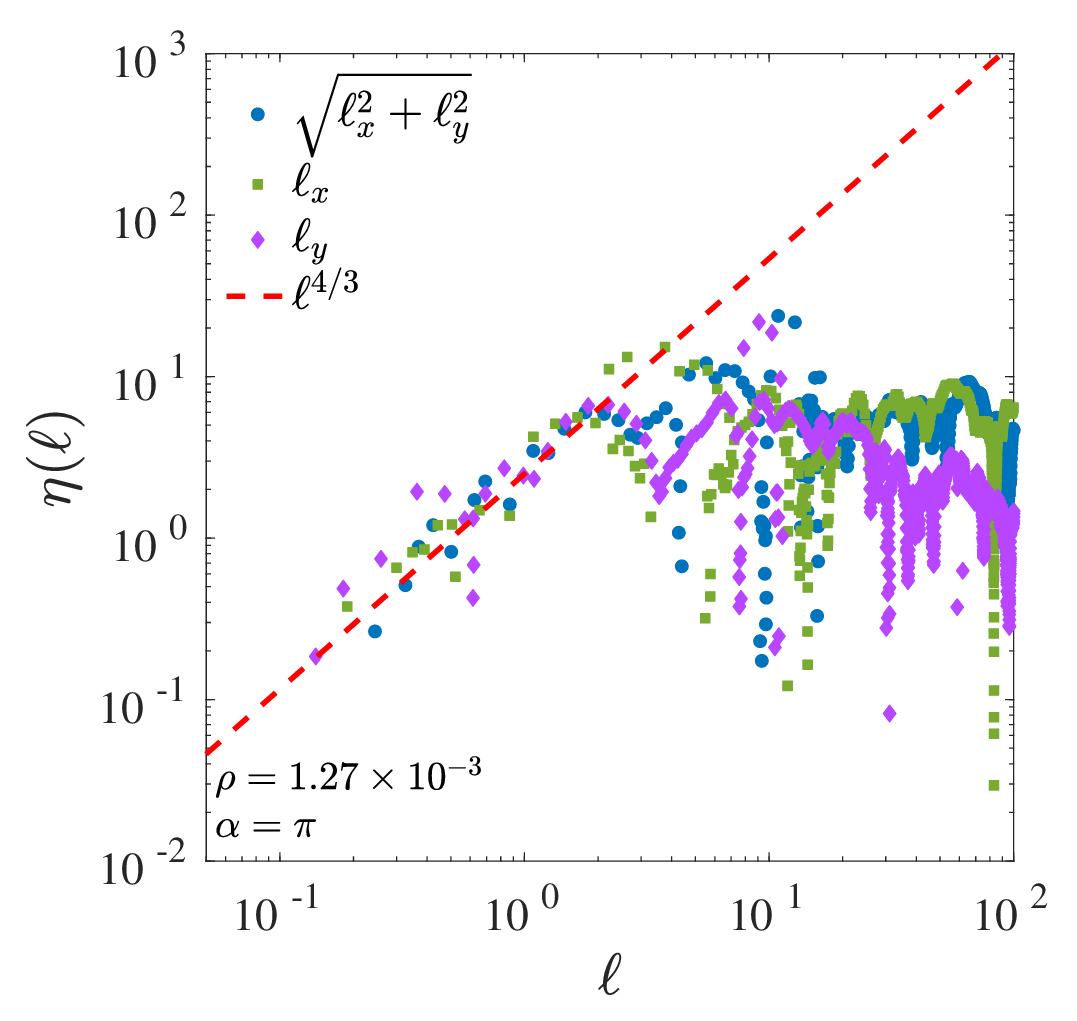}
	\caption{Scaling of the eddy diffusivity $\eta(\ell)$ for the various components of the tensor $\sigma_ij$, for the isotropic case $E$. The red dashed line represents the power scaling $\eta\sim\ell^{4/3}$, the vertical dashed line represents the forcing scale $\ell_c = 2\pi/k_c$.}
	\label{fig:kappa} 
\end{figure}

Finally, in Figure~\ref{fig:kappa}, is reported the scaling of the eddy diffusivity $\eta(\ell)$ for the isotropic case $E$. A power-law scaling compatible with the classical Richardson's prediction, $\eta\sim\ell^{4/3}$, is observed (indicated by the red dashed line).

\begin{figure}[h]
	\centering\includegraphics[scale=0.4]{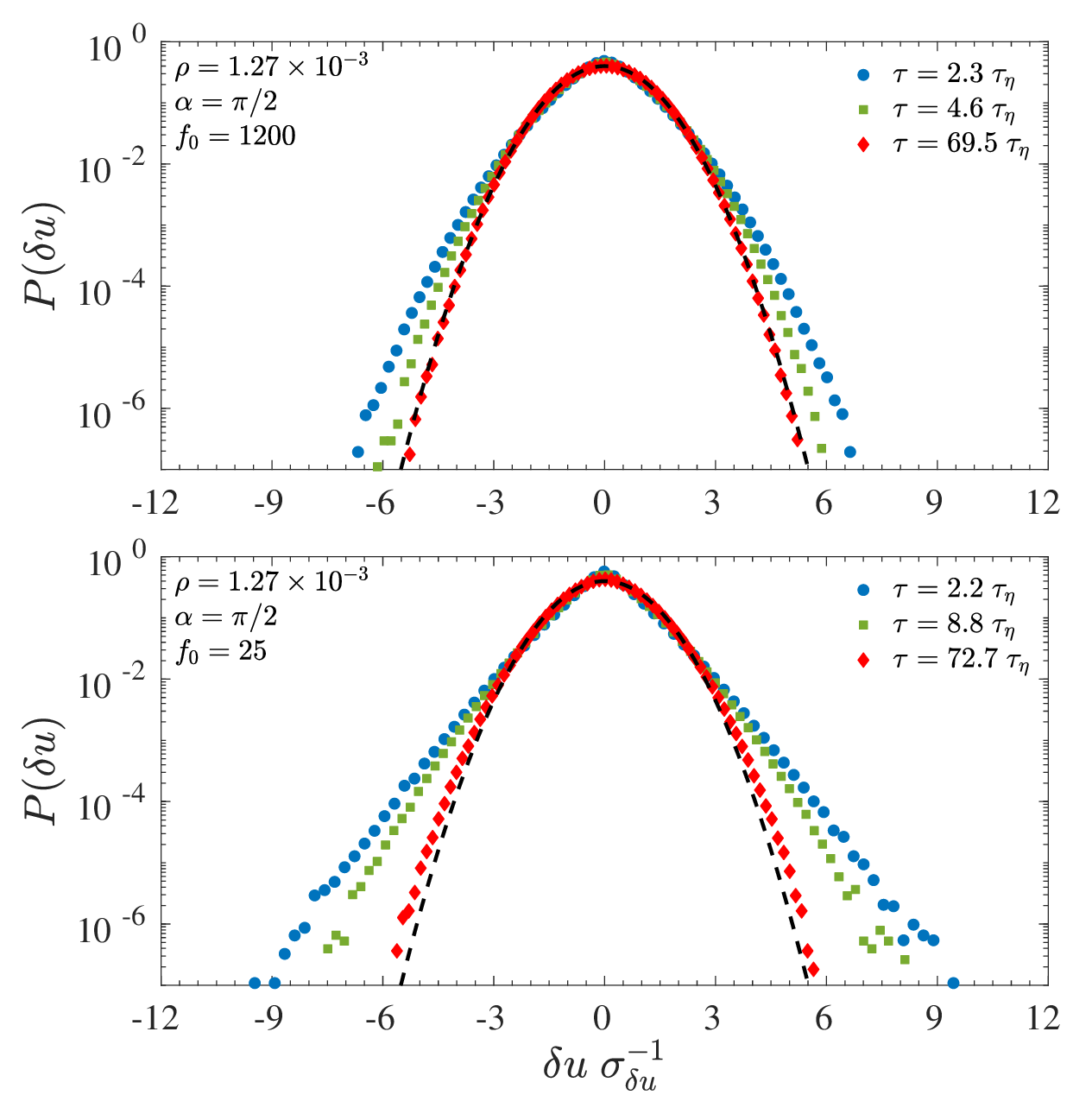}
	\caption{
		PDFs of Lagrangian velocity fluctuations ($\delta u$), normalized by their root-mean-square ($\sigma_{\delta u}$) value, for various lag times $\tau$, under two different forcing conditions. The top panel reports the PDF for a system with a forcing frequency of $f_0=1200$, corresponding to a low energy injection rate and weak intermittency. The bottom panel presents the PDF for a system with $f_0=25$, indicative of a higher energy injection rate and strong intermittency. In both panels, the Gaussian distribution is shown as a dashed line for visual comparison.}
	\label{fig:PDF} 
\end{figure}

As energy injection rate $\epsilon_f$ is increased, the system leads to stronger intermittency effects in the Lagrangian velocity fluctuations $\delta u = u(t+\tau) - u(t)$. This is a fundamental characteristic of fully developed turbulence, where energy is not dissipated uniformly but is concentrated into rare, intense events.
This phenomenon is most clearly observed by analyzing the Probability Density Functions (PDFs) of the Lagrangian velocity fluctuations at small scales (Figure~\ref{fig:PDF}, being $\tau_\eta$ the Kolmogorov time-scale).
The tails of the PDFs grow significantly when the forcing amplitude is modified, hence the injection rate, deviating sharply from a normal distribution. This indicates that the probability of observing large, extreme velocity fluctuations becomes much higher than what a simple random process would predict.

The growth of these non-Gaussian tails is the hallmark of the multifractal nature of the energy dissipation field. In a highly turbulent flow, the dissipation of enstrophy is not uniform throughout the fluid. Instead, it is concentrated in intricate, fractal-like structures such as thin filaments and sheets.

\section{Conclusions}

The challenge of modeling turbulence at high Reynolds numbers has driven research toward the adoption of innovative simulation methods. In this context, a different approach was introduced, which is based on a truncated 2D Navier-Stokes model with a randomized and pseudo-logarithmic grid in Fourier space. This unconventional methodology provided a controlled and computationally efficient environment for a deep investigation into the influence of anisotropy (or non-uniform mode distribution) and the interacting triads density on turbulence dynamics and Lagrangian statistics. Through this model, not only was the reproduction of classical 2D turbulence phenomena confirmed, but new and significant insights were also gained into how the structure of Fourier space is reflected in the behavior of the real flow.

The primary findings of this study are summarized below:

	(i) The influence of anisotropy, which was introduced by sampling specific angular sectors of Fourier space, was effectively captured. The vorticity fields in anisotropic cases are characterized by elongated, filamentary structures and a preferential direction for energy transfer, which is in sharp contrast with the homogeneous and finely-structured fields of the isotropic cases. Crucially, it was found that these large-scale anisotropic effects do not prevent the emergence of the classical spectral scaling laws, although the efficiency of the inverse energy transfer was observed to be slightly reduced.

	(ii) The double cascade scenario characteristic of 2D turbulence was successfully reproduced by the model. Energy, which was injected at the forcing scale, was observed to exhibit an inverse cascade towards large scales, with an energy spectrum compatible with the theoretical prediction of $E(k) \sim k^{-5/3}$.
	This was further confirmed by the sign of the energy flux, $\Pi_E(k)$, which was found to be negative in the inverse cascade range. Simultaneously, enstrophy was transferred to small scales via a direct cascade, where it was dissipated by viscosity, leading to a steeper spectral scaling of $E(k) \sim k^{-3}$. It was confirmed that these fundamental energy and enstrophy transfer mechanisms are correctly captured by the model's low-density Fourier space, despite its non-uniformity.
	
    (iii) A {Richardson-like super-ballistic diffusion regime} for particle pairs was revealed by the analysis of particle dynamics, with the mean squared separation scaling as {$\ell^2 \sim t^3$}. This scaling was observed to be robust across different initial separations ($R_0$) and is driven by a trapping-and-ejection mechanism from vortices, which was shown to be a fundamental feature of the model's voricity field. It was also demonstrated that the eddy diffusivity $\eta(\ell)$ follows the predicted {$\eta \sim \ell^{4/3}$} scaling, in a range of $\ell > 2\pi k_c^{-1}$. For longer timescales, a transition to a {Brownian-like diffusion regime} ($\ell^2 \sim t$) was observed.
{The emergence of a $t^3$ pair-dispersion indicates that the retained spectral interactions are sufficient to generate velocity fluctuations supporting super-ballistic separation, and
    anisotropy mainly affects the dispersion amplitude, while the scaling itself remains robust, suggesting that it is governed by the spectral organization of the nonlinear interactions rather than by geometric isotropy.}	
	
	(iv) Strong evidence for the phenomenon of intermittency was provided by the model. By varying the energy injection rate $\epsilon_f$, the Probability Density Functions (PDFs) of Lagrangian velocity fluctuations were observed to develop heavy, non-Gaussian tails. This is considered a key signature of fully developed turbulence, where energy dissipation is not uniform but is concentrated in rare, intense events.

The ability of this truncated system to reproduce such complex statistical behavior highlights a crucial link between the geometry of the Fourier mode distribution and the multifractal nature of turbulent dissipation in real space. Future studies are needed to investigate these properties in greater detail, specifically to gain a deeper understanding of how the system's multifractal characteristics relate to the underlying flow dynamics and the interacting triadic density.

\begin{acknowledgements}
	F.C. acknowledges the contribution received from program Fondo per il
	Programma Nazionale di Ricerca e Progetti di Rilevante Interesse Nazionale
	(PRIN) under the project ``TURBIMECs: study of TURBulence In MEditerranean Cyclone events'', grant n. 2022S3RSCT, CUP Master B53D23007500006, funded by the Italian Ministry of University and Research (MUR).	S. S. acknowledges the Space It Up project funded by the Italian Space Agency, ASI, and the Ministry of University and Research, MUR, under Contract No. 2024-5-E.0-CUP No. I53D24000060005.
\end{acknowledgements}

%\newpage
%\bibliography{V29}% Produces the bibliography via BibTeX.

%apsrev4-2.bst 2019-01-14 (MD) hand-edited version of apsrev4-1.bst
%Control: key (0)
%Control: author (8) initials jnrlst
%Control: editor formatted (1) identically to author
%Control: production of article title (0) allowed
%Control: page (0) single
%Control: year (1) truncated
%Control: production of eprint (0) enabled
%

\end{document}